\documentclass[conference]{IEEEtran}
\IEEEoverridecommandlockouts
\usepackage{url}
\usepackage{cite}
\usepackage{amsmath,amssymb,amsfonts}
\usepackage{algorithmic}
\usepackage{graphicx}
\usepackage{array}
\usepackage{textcomp}
\usepackage{hyperref} 
\usepackage[utf8]{inputenc}
\usepackage{booktabs}
\usepackage{tabularx}
\usepackage{xcolor}
\usepackage{colortbl}
\def\BibTeX{{\rm B\kern-.05em{\sc i\kern-.025em b}\kern-.08em
    T\kern-.1667em\lower.7ex\hbox{E}\kern-.125emX}}
\begin{document}

\title{Developments in Building Towards the Indian SKA Regional Centre}

\author{
  \IEEEauthorblockN{Rohit Sharma}
  \IEEEauthorblockA{\textit{Department of Space, Planetary \& Astronomical} \\ 
  \textit{Sciences \& Engineering (SPASE)} \\
  \textit{Indian Institute of Technology Kanpur}\\
  \textit{Kalyanpur, Kanpur, 208016, U.P., India} \\
  rsharma@iitk.ac.in \\ ORCID: \url{https://orcid.org/0000-0003-0485-7098}}
  \and
  \IEEEauthorblockN{Yogesh Wadadekar}
  \IEEEauthorblockA{\textit{National Centre for Radio Astrophysics,} \\ 
  \textit{Pune University Campus, Ganeshkhind,} \\ 
  \textit{Pune - 411007, Maharashtra, India}\\
  yogesh@ncra.tifr.res.in \\ ORCID: \url{https://orcid.org/0000-0002-1345-7371}}
}

\maketitle 

\begin{abstract}
The SKA Regional Centre Network (SRCNet) is a federated global network designed to archive, distribute, and process SKAO data products. This paper examines the evolving SRCNet architecture-focusing on the transition to version 0.2 (v0.2) requirements, including computation, orchestration, and multi-tiered storage-and details the strategy for establishing its proto-SRC infrastructure. We outline the prototype developments for the major and minor nodes in the context of v0.2. We describe the minor node compute and storage demands, establishing a pathway toward a fully operational Indian SKA Regional Centre (INSRC).

\end{abstract}

\begin{IEEEkeywords}
Radio astronomy, High-performance computing, Distributed computing, Computer architecture, Software systems
\end{IEEEkeywords}

\section{Introduction}

The under-construction Square Kilometre Array Observatory (SKAO) will produce high-quality radio data products. The sheer volume of data produced by the SKAO (roughly 700 Petabytes per year) cannot be handled by a single facility \cite{b2,b3}. Instead, SKAO will rely on the SKA Regional Centre Network (SRCNet)—a federated, global system of computing centres aimed at archiving, distributing, and processing the radio data products based on computing capacity and user-specific requirements \cite{b4}. The SRCNet will enable SKA science users to analyse complex data products. It will serve as the primary portal for researchers to access SKAO data, offer resources for analysis and visualisation, and drive innovation in distributed computing and science gateways. 

\subsection{Development Phases of SRNet}
SRCNet development follows an iterative roadmap, progressing from early internal releases (such as v0.1 and v0.2) aimed at establishing foundational site interconnections, basic data replication, and remote workflow execution, before scaling into full operational capability to support global science processing.

To support the scaling of processing pipelines and data management workflows within the SRCNet, the underlying infrastructure transitions across multiple operational phases: v0.1 (foundational deployment), v0.2 (scaled operational deployment), v0.3 and so on \footnote{ \href{https://docs.google.com/document/d/1PZ4Il_RgIs2rtR0XawoAa0Q3FXycI4-4yhmjbbrDzLw/edit?tab=t.0}{SRC-0000008-01, SRCNet v0.1 Node Requirements}}. The SRCNet v0.1 is already operational, while v0.2 starts in 2027. The major requirement transition from v0.1 to v0.2 is listed below.

\begin{itemize}
    \item Compute Infrastructure
    \begin{itemize}
        \item Version v0.1: Focuses on establishing a baseline compute framework built around CPU and virtual machine (VM) environments. It utilises standard batch scheduling mechanisms to handle preliminary data processing tasks.
        \item Version v0.2: Scales up to a heterogeneous compute architecture incorporating GPU acceleration to execute computationally intensive data pipelines. Workload management transitions to a dynamic, Kubernetes-based orchestration framework to support containerised, highly scalable processing workflows.
    \end{itemize}
        \item Storage Architecture
    \begin{itemize}
        \item Version v0.1: Employs static storage endpoints configured for fundamental data movement and simple replication across nodes using Rucio.
        \item Version v0.2: Expands to a multi-tiered, dynamic storage system featuring automated data lifecycle policies. This configuration enables intelligent caching, automated staging, and seamless data movement between high-performance local scratch storage and long-term archival tiers.
    \end{itemize}
        \item Network Infrastructure
    \begin{itemize}
        \item Version v0.1: Relies on standard academic backbones operating at 10–100 Gbps speeds with conventional routing protocols to facilitate initial node interconnectivity.
        \item Version v0.2: Introduces optimised network routing alongside dedicated bandwidth reservations to guarantee high-throughput, low-latency data transfers across geographically distributed nodes during peak pipeline execution.
    \end{itemize}
\end{itemize}

Under the v0.2 deployment architecture, accredited nodes run standardised Mandatory Local Services (including Rucio, JupyterHub, metadata servers, and visualisation containers) alongside Optional Local Extensions tailored for VO services or HPC integration. These distributed components are interconnected through centralised Global Services, which manage federated execution, global metadata, network monitoring, and unified Identity and Access Management (IAM).

Acting as both a foundational stepping stone toward early science verification and the primary data computing gateway for SKAO users worldwide, SRCNet v0.2 represents a key phase in the global deployment roadmap. For the v0.2 deployment, the global requirement estimates are based on an eight-node network topology comprising three major nodes and five minor nodes. In this paper, we discuss the requirements and challenges in the context of the developing SRCNet deliveries.

\subsection{Worldwide Participation}

SRCNet v0.1 served as a proof-of-concept for the global architecture, validating core service deployments and inter-node connections while gathering critical feedback from operators and scientists to guide future development. Running software version 0.1, the system has successfully met all required deliverables, with nine regional centres—Spain, Switzerland, the SKAO, Sweden, the UK, China, Japan, Canada, and Italy—passing the official accreditation test.


\section{Ongoing Indian Proto-SRC Efforts}

\begin{figure}[htbp]
\centerline{\includegraphics[width=0.5\textwidth]{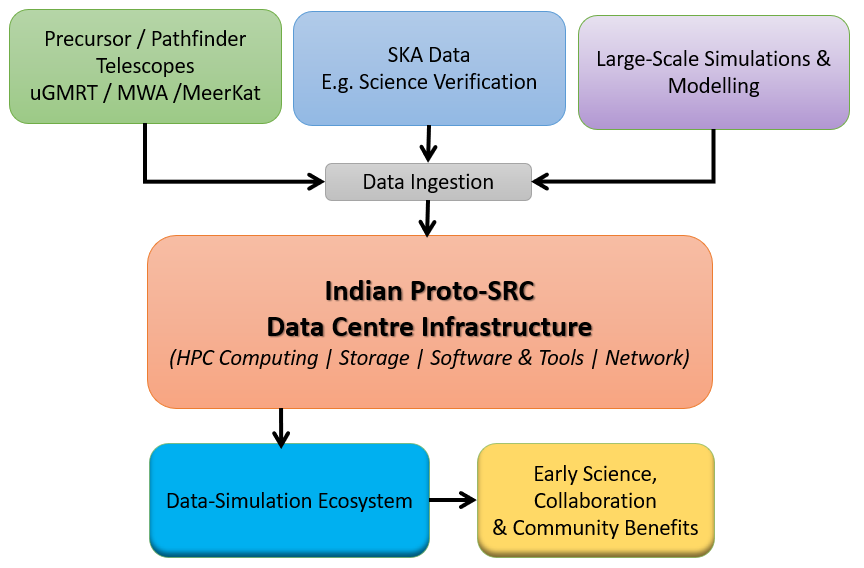}}
\caption{Indian proto-SRC development \& usage.}
\label{fig}
\end{figure}

In alignment with global SRCNet developments, India proposed establishing a proto-SRC stationed at NCRA-TIFR, Pune \cite{b1}. Under construction, this facility is aimed at evaluating emerging hardware and software architectures while driving targeted software development in key SKA data processing domains, including algorithm execution, pipeline optimisation, and data visualisation tools. The usage of Indian proto-SRC has been envisaged in three major modes: (1) process data from SKA precursors \& pathfinders, (2) SKA data processing, e.g. for science verification, etc., and (3) supporting the observations with large-scale simulations. Fig. \ref{fig} shows the data ingestion from these modes to the centre infrastructure, leading to a coherent data-simulation ecosystem and ultimately community benefits.

\subsection{Compute Infrastructure}

For the proto-SRC, the plan is to deploy server-class CPU compute hardware providing roughly 20 TFLOPS of processing power, supplemented by GPU acceleration cards on targeted servers. The precise CPU-to-GPU capacity balance will be optimised by benchmarking actual radio astronomy data workloads. It is designed for a five-year operational lifecycle; this compute stack will provide baseline processing capability before being superseded by full SRC hardware. In addition to the major node, the minor nodes are being set up at the Indian Institute of Technology, Kanpur (IITK), and the Indian Institute of Science, Bangalore (IISc). These minor nodes are being set up under the infrastructure of the National Supercomputing Mission (NSM).

\subsubsection{Activities at NCRA-TIFR}

A small prototype has been set up at NCRA-TIFR. Currently, this prototype node setup has three primary functional layers:
\begin{itemize}
    \item Physical / Virtualisation Layer: 10 bare-metal physical nodes running Proxmox VE 8.2 for high-performance virtualisation.
    \item Identity \& Subnet Services: Primary and replicated FreeIPA virtual machine providing failover identity management and subnet-wide DNS.
    \item Kubernetes Orchestration Layer: It hosts NGINX load balancing, 3 control plane nodes, and 10 distributed worker nodes deployed over AlmaLinux 9.4 virtual machines and configured to distribute API requests seamlessly.
\end{itemize}

To ensure high availability of this prototype, virtual machines are distributed across the 10-node Proxmox cluster in a fault-tolerant layout. The management operations and administrative workflows are run from a dedicated host and have been optimised for resource usage. The plan is to test this prototype in SRCNet v0.2. This under-development prototype will also act as a launchpad to establish minor nodes infrastructure.

\begin{table*}[!htbp]
\centering
\small
\begin{tabularx}{\textwidth}{l X X X X}
\toprule
\textbf{Resource} & \textbf{SRCNet v0.2 Minor Node} & \textbf{Upgraded Param} & \textbf{Existing Param (HM*)} & \textbf{Remarks} \\
\midrule
\textbf{Compute} 
  & 0.4 PFlops 
  & 2.5 PFlops 
  & 1.67 PFlops 
  & 200 GFlops per core for SRCNet; Upgraded Param will have 2.5 PF CPU and 2.5 PF GPU compute \\
\midrule
\textbf{No. of Cores} 
  & 2000 
  & -- 
  & 3744 
  & 78 HM CPU with 48 cores each; Upgraded Param has Intel Xeon Scalable processor with 64 cores \\
\midrule
\textbf{Storage} 
  & 2.4 PB 
  & 5.0 PB 
  & 1.1 PB 
  & -- \\
\midrule
\textbf{Scratch/Cache} 
  & 300 TB -- 400 TB 
  & -- 
  & 480 GB per node 
  & -- \\
\midrule
\textbf{RAM per node} 
  & 384 GB 
  & 4 TB 
  & 768 GB 
  & -- \\
\bottomrule
\end{tabularx}
\caption{Summary table for the SRCNet v0.2 requirement and IITK's NSM Infrastructure (*HM: High Memory CPU)}
\label{tab:srcnet_v02_nsm_iitk}
\end{table*}

\subsection{Storage Infrastructure}

The target infrastructure includes ~10 PB of storage and balanced compute resources to drive data analysis and simulation workloads. The storage stack will leverage a hybrid HDD/SSD architecture balanced for cost and performance, with NVMe options evaluated to handle high-speed data access.

\subsection{Networking Challenges}
To support offsite users distributed across India, high-speed connectivity for the proto-SRC will be delivered via the National Knowledge Network (NKN)—the country’s multi-gigabit NREN connecting thousands of research and academic institutions all over India. NKN’s scalable bandwidth will satisfy the networking requirements of SKA India Consortium members, though select institutions may require connection upgrades to handle the increased usage.

\subsection{Minor node plans at IIT Kanpur}

The minor nodes have relatively smaller requirements for SRCNet. Table \ref{tab:srcnet_v02_nsm_iitk} lists requirements from minor nodes for SRCNet v0.2. For Indian proto-SRC, this will be set up under NSM facilities. IIT Kanpur hosts the Param Sanganak HPC facility to support multidomain research. Computational capacity on Param Sanganak (1.67 PFLOPS) is distributed under a fair-share allocation model, reserving up to 40\% of compute resources for non-IIT Kanpur external users. In 2026, a major upgrade is planned for the NSM facility. For convenience in this note, we call it the upgraded-param (uParam). Table \ref{tab:srcnet_v02_nsm_iitk} also lists resources from existing Param and uParam. Both Param Sanganak and the upcoming NSM facility at IIT Kanpur comfortably meet the baseline minor node specifications for 2026. Deploying SRCNet v0.2 requires 0.4 PFLOPS of compute (calculated at 200 GFLOPS per core across 2,000 required cores) to support federated science workflows and SKA-scale data processing. With the uParam offering a total capacity of 5 PFLOPS split equally between CPU and GPU architectures, the SRCNet v0.2 requirement represents only about 16\% of the overall system.

\section{Future Directions}

Ongoing work to set up an Indian proto SRC is a crucial initial step for a larger Indian SKA Regional Centre (INSRC). The services across computational and data infrastructures are designed to support the Indian user community to deliver SKA science, under the FAIR (Findable, Accessible, Interoperable, Reusable) principles. Ongoing developments of the prototype node at NCRA-TIFR and the upcoming minor at IIT Kanpur will reinforce the proto-SRC setup in the country. Construction of the full INSRC needs to be planned for the coming years, centred around a dedicated data centre facility engineered to handle data streams from both SKA telescopes. In close alignment with global SRCNet partners, the development timeline will ensure the infrastructure is fully operational to support SKA science verification and long-term operations by the end of the decade.

\section*{Acknowledgment}

RS acknowledge Saurabh Mookherjee at Thoughtworks, SRCNet prototype teams- Teal \& project scientists, and Swiss-SRC colleagues for fruitful discussions and clarifications. RS also acknowledges IIT Kanpur for continuing support for SKA-related activities. YW acknowledges support from the Department of Atomic Energy, Government of India under project UID number RTI4017.

\end{document}